\documentclass[twocolumn,aps,prl,showpacs,amsmath,amssymb,floatfix,superscriptaddress]{revtex4-1}
\usepackage{dcolumn}
\usepackage{bm}
\usepackage{amsmath}
\usepackage{amsfonts}
\usepackage{graphicx}
\usepackage{url}
\usepackage{hyperref}
\usepackage{color}
\usepackage{float}
\begin{document}
\title{Survival of the Scarcer}
\author{Alan Gabel}
\affiliation{Center for Polymer Studies and Department of Physics, Boston University, Boston,
MA 02215, USA}
\author{Baruch Meerson}
\affiliation{Racah Institute of Physics, Hebrew University, Jerusalem 91904, Israel}
\author{S. Redner}
\affiliation{Center for Polymer Studies and Department of Physics, Boston University, Boston,
MA 02215, USA}

\begin{abstract}

  We investigate extinction dynamics in the paradigmatic model of two
  competing species $A$ and $B$ that reproduce ($A\to 2A$, $B\to 2B$),
  self-regulate by annihilation ($2A\to 0$, $2B\to 0$), and compete ($A+B\to
  A$, $A+B\to B$).  For a finite system that is in the well-mixed limit, a
  quasi-stationary state arises which describes coexistence of the two
  species.  Because of discrete noise, both species eventually become extinct
  in time that is exponentially long in the quasi-stationary population size.
  For a sizable range of asymmetries in the growth and competition rates, the
  paradoxical situation arises in which the numerically disadvantaged species
  according to the deterministic rate equations survives much longer.

\end{abstract}
\maketitle

In the paradigmatic two-species competition model, a population is comprised
of distinct species $A$ and $B$, each of which reproduce and self regulate by
intraspecies competitive reactions.  In addition, \emph{interspecies}
competitive reactions occur, which are deleterious to both species~\cite{M}.
For large, well-mixed populations, the dynamics can be accurately described
by deterministic rate equations.  For finite systems, however, fluctuations
in the numbers of individuals ultimately lead to extinction, in stark
contrast to the rate equation predictions.

In this work, we investigate how \emph{asymmetric} interspecies competition
influences the extinction probability of each species.  In a finite
ecosystem, extinction arises naturally when multiple species compete for the
same resources.  In such an environment, one species often dominates, while
the others become extinct~\cite{H60,F88,B89,NF11,G11}, a feature that
embodies the {\it competitive exclusion principle}.  A related paradigm
appears in the context of competing parasite strains that exploit the same
host population, or in the fixation of a new mutant allele in a haploid
population whose size is not fixed \cite{Parsons}.

With asymmetric interspecies competition, we uncover the surprising feature
that deterministic and stochastic effects, which originate from the same
elemental reactions, act oppositely.  For sizable asymmetry ranges in the
growth and competition rates, the situation arises where the combined effects
of the elemental reactions leads to one species being numerically
disadvantaged at the mean-field level, \emph{despite} its interspecies
competitive advantage, but this competitive advantage dominates the other
reaction processes at the level of large deviations.  Thus the outcompeted
and less abundant species has a higher long-term survival probability:
``survival of the scarcer''.

\medskip
\noindent{\it Model:} Asymmetric competition of two species A and B is
defined by the reactions:
\begin{eqnarray}
\label{reactions}
&~~~A  \stackrel{1}{\longrightarrow}  A+A\qquad\qquad
&B  \stackrel{g}{\longrightarrow}  B+B\,,\nonumber\\
&~~~A+A  \stackrel{1/K}{\longrightarrow}  0\qquad\qquad
&B+B  \stackrel{1/K}{\longrightarrow}  0\,,\\
&~~~A+B  \stackrel{\epsilon/K}{\longrightarrow}  B \qquad\qquad
&A+B  \stackrel{\alpha\epsilon/K}{\longrightarrow}  A.\nonumber
\end{eqnarray}
The first line accounts for reproduction, the second for intraspecies
competition, and the last for interspecies competition.  Here $K$ is the
environmental carrying capacity, which sets the size of the overall
population, $\epsilon$ quantifies the severity of the competition, while $g$
and $\alpha$ quantify the asymmetries in the growth and interspecies
competition rates, respectively.  In our presentation, we focus on the limit
$K\gg 1$.  While a general model should also contain asymmetry in the
intraspecies competition rate, no new phenomena arise by this generalization;
for simplicity, we study the model defined by Eqs.~\eqref{reactions}.

To probe extinction in two-species competition, we focus on $P_{m,n}(t)$, the
probability that the population consists of $m\geq 0$ As and $n\geq 0$ Bs at
time $t$.  In the limit of a perfectly-mixed population, the stochastic reaction
processes in \eqref{reactions} lead to $P_{m,n}(t)$ evolving by the master
equation
\begin{align}
\label{master}
\dot{P}_{m,n}(t)&=\hat{H} P_{m,n} = \big[\!\left( \mathbb{E}^{-1}\!-\!1 \right)m
+g\left( \mathbb{F}^{-1}\!-\!1\right)n\big]P_{m,n} \nonumber\\
& +\Big[\!\left( \mathbb{E}^{2}\!-\!1 \right)\frac{m(m\!-\!1)}{2K}
+\left( \mathbb{F}^{2}\!-\!1 \right)\frac{n(n\!-\!1)}{2K}\Big]P_{m,n} \nonumber\\
& +\left[\frac{\epsilon}{K}\left( \mathbb{E}-1 \right)
+\frac{\alpha\epsilon}{K}\left( \mathbb{F}-1\right)\right]mnP_{m,n}~.
\end{align}
Here $\mathbb{E}$ and $\mathbb{F}$ are the raising and lowering operators~\cite{V01} for
species A and B, respectively; viz.\ $\mathbb{E}^{i}P_{m,n}=P_{m+i,n}$ and
$\mathbb{F}^{j}P_{m,n}=P_{m,n+j}$.

\begin{figure}[ht]
  \begin{center}
    \includegraphics[width=0.2\textwidth]{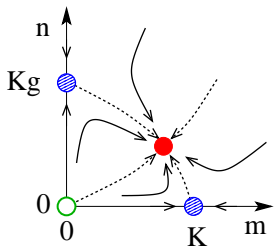}
    \caption{\small Schematic flow diagram in asymmetric two-species
      competition for weak competition in the mean field.  The unstable node,
      the saddles, and the stable node are shown as open, hatched, and solid,
      respectively. }
 \label{flow-mf}
  \end{center}
\end{figure}

\medskip
\noindent{\it Deterministic Rate Equations:}
First we focus on the average population sizes $\langle m\rangle =\sum_{m,n}mP_{m,n}$
and $\langle n\rangle =\sum_{m,n}nP_{m,n}$.  From \eqref{master}, the
evolution of these quantities is given by
\begin{align}
\begin{split}
\label{meanField}
\dot{\langle m\rangle}&= \langle m\rangle\left( 1-\frac{\langle m\rangle}{K}
-\epsilon\frac{\langle n\rangle}{K} \right)\,, \\
\dot{\langle n\rangle}&= \langle n\rangle\left( g-\frac{\langle n\rangle}{K}
-\alpha\epsilon\frac{\langle m\rangle}{K} \right)\,.
\end{split}
\end{align}
Here we neglect correction terms of the order of $1/K$ and, more importantly,
neglect correlations by assuming that $\langle m^2\rangle=\langle
m\rangle^2$, $\langle n^2\rangle=\langle n\rangle^2$, and $\langle
mn\rangle=\langle m\rangle \langle n\rangle$.  We restrict ourselves to the
parameter range $\alpha\epsilon<g<1/\epsilon$, which guarantees that the
fixed point corresponding to coexistence of both species is stable.  The four
fixed points of the rate equations \eqref{meanField} are then:
\begin{align}
\label{FP}
(m^*, n^*)&= (0,0)\quad\qquad \qquad\qquad\mathrm{unstable\ node},\nonumber\\
 &= (K,0)\,, (0,Kg)~~~\qquad \mathrm{saddles},\\
 &=
 \left(K\tfrac{1\!-\!g\epsilon}{1\!-\!\alpha\epsilon^2},K\tfrac{g\!-\!\alpha\epsilon}{1\!-\!\alpha\epsilon^2}
 \right)~\quad\mathrm{stable\ node}.\nonumber
\end{align}
If the initial populations of both species are non-zero, they are quickly
driven to the stable node (Fig.~\ref{flow-mf}) that describes the
steady-state populations in the mean-field limit.  The relaxation time toward
the stable node, $\tau_r$, is independent of $K$. These steady-state
populations of the two species are equal when
\begin{equation}\label{equalsize}
g^* =\frac{1+\alpha \epsilon}{1+\epsilon}.
\end{equation}
For $g<g^*$, the $B$-population is scarcer.  Naively, the scarcer population
should be more likely become extinct first.  However, as we shall show, a
proper account of the fluctuations that stem from the underlying elemental
reactions themselves leads to a radically different outcome.

\noindent{\it Extinction:} The mean-field picture is incomplete because
fluctuations of the population sizes about their fixed-point values are
ignored.  For large populations (corresponding to carrying capacity $K\gg
1$), these fluctuations are typically small.  Thus the populations achieve a
\emph{quasi-stationary} state where the two species coexist.  This state is
stable in the mean-field description (heavy dot in Fig.~\ref{flow-mf}).
However, an unlikely sequence of deleterious events eventually occurs that
ultimately leads one population, and then the other, to extinction.  After
the first extinction, the remaining population settles into another
quasi-stationary state around one of the single-species fixed points
$(m^*,n^*)=(K,0)$ or $(0,Kg)$.  Eventually a large fluctuation drives the
remaining species to extinction.  This second extinction time is typically
much longer [by a factor that scales as $\exp(\text{const.}\times K$)] than
the first time, because the remaining species does not suffer interspecies
competition.  Once a species is extinct, there is no possibility of recovery
since there is no replenishment mechanism.

\begin{figure}[ht]
\center{
\includegraphics[width=0.45\textwidth]{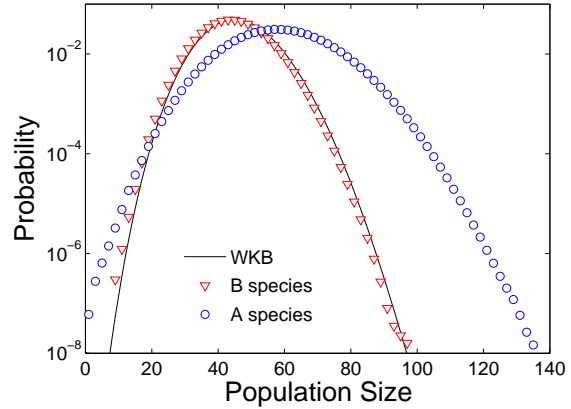}
\caption{Quasi-stationary probability distributions for species A, $P_m=\sum_{n} P_{m,n}$, and species B, $P_n=\sum_{m} P_{m,n}$.  Parameters are $K=100$, $\epsilon=0.9$, $g=0.45$, and
  $\alpha=0$.  Symbols are simulation results, while the solid
  curve is WKB approximation for the B species
  distribution.}
\label{distributions}
}
\end{figure}

The question that we address is: which species typically goes extinct first?
The answer is encoded in the dynamics of the two-species probability
$P_{m,n}(t)$.  During the initial relaxation stage, a quasi-stationary
probability distribution is quickly reached (Fig.~\ref{distributions}).  The
probability distribution is sharply peaked at the stable fixed point of the
mean-field theory.  This probability slowly ``leaks" into localized regions
near each of the single-species fixed points $(m^*,n^*)=(K,0)$ and $(0,Kg)$.
Thus two sharply-peaked single-species distributions start to form.  If the
$(K,0)$ peak grows faster then the B species is more likely to go extinct
first.  Similarly, a faster growing $(0,Kg)$ peak means A extinction is more
likely.  Eventually, the probability distribution that is localized at one of
the two single-species fixed points slowly leaks toward the fixed point
$(0,0)$ that corresponds to complete extinction~\cite{movie}.

To determine extinction rates, it is helpful to define $\mathcal{P}_A$,
$\mathcal{P}_B$, and $\mathcal{P}_\phi$ as the respective probabilities that
species A is extinct, species B is extinct, or neither is extinct at time $t$
\cite{GM12}. (Being interested in times much shorter than the expected
extinction time of \emph{both} species, we can neglect the probability of the
latter process.) By definition, these extinction probabilities are
\begin{eqnarray}
\label{coarseDefn}
\mathcal{P}_A\!=\! \sum_{n>0}\!P_{0,n}, \quad
\mathcal{P}_B\!=\! \sum_{m>0}\!\!P_{m,0},\quad
\mathcal{P}_\phi\!=\!\!\! \sum_{m,n>0}\!\!\!P_{m,n};
\end{eqnarray}
these satisfy $\mathcal{P}_A+\mathcal{P}_B+\mathcal{P}_\phi= 1$, up to an
exponentially small correction that stems from the process where both species
become extinct simultaneously.  In the limit $K\gg 1$ and for times much
greater than the relaxation time scale $\tau_r$, the sums in
Eqs.~\eqref{coarseDefn} are dominated by contributions from values of $m$ and
$n$ that are close to the single-species and coexistence fixed points.
Moreover, these extinction probabilities evolve according to a set of
effective coupled equations
\begin{align}
\begin{split}
\label{coarseMaster}
\dot{\mathcal{P}}_{A}&=R_A\mathcal{P}_\phi\,, \\
\dot{\mathcal{P}}_{B}&=R_B\mathcal{P}_\phi\,,\\
\dot{\mathcal{P}}_\phi&=-(R_A+R_B)\mathcal{P}_\phi\,,
\end{split}
\end{align}
that define $R_A$ and $R_B$ as the respective extinction rates for species A
and species B.  Solving these equations yields the time dependence of the
extinction probabilities
\begin{equation}
\label{extinction}
\mathcal{P}_A(t)= \frac{R_A}{\mathcal{R}}\big(1\!-\!e^{-\mathcal{R}t}\big)\,,\quad
\mathcal{P}_B(t)= \frac{R_B}{\mathcal{R}}\big(1\!-\!e^{-\mathcal{R}t}\big)\,,
\end{equation}
with $\mathcal{R}=R_A+R_B$. To determine $R_A$ and $R_B$, we follow the
evolution of the eigenstate of the master equation (\ref{master}) that
determines the leakage of probability from the vicinity of the coexistence
point:
\begin{equation}
\label{quasiSteady}
P_{m,n}(t)= \Pi_{m,n}\,e^{-\mathcal{R} t},\qquad  m,n>0,
\end{equation}
where
\begin{equation}\label{eigen}
    \hat{H} \,\Pi_{m,n} = -\mathcal{R}\, \Pi_{m,n}, \qquad  m,n>0,
\end{equation}
and $\mathcal{R}$ is the \emph{third}-lowest positive non-trivial eigenvalue
of the operator $\hat{H}$. The two still-smaller positive non-trivial
eigenvalues correspond to the much slower decay of quasi-stationary
single-species states and play no role in the dynamics of the first
extinction event.  There is also a trivial eigenvalue that corresponds to the
final state of complete extinction.

Combining Eq.~\eqref{master} with \eqref{coarseDefn}--\eqref{quasiSteady},
we obtain the following expression for the extinction rate of the A species:
\begin{subequations}
\begin{equation}
  R_A=\frac{1}{K}\sum_{n>0}\big(\epsilon\, n\, \Pi_{1,n}+\Pi_{2,n}\big)\,.
\label{rate1}
\end{equation}
As expected, the extinction rate for As involves two processes: (i)
elimination of the last remaining A via competition with Bs and (ii)
annihilation of the last remaining pair of As.  Similarly,
\begin{equation}
  R_B=\frac{1}{K}\sum_{m>0}\big(\alpha\epsilon\, m\,\Pi_{m,1}+\Pi_{m,2}\big)\,.
\label{rateB}
\end{equation}
\end{subequations}
To calculate $R_A$ and $R_B$, we therefore need to evaluate the
small-population-size  tails of $\Pi_{m,n}$.  This task can be achieved by
applying a variant of Wentzel-Kramers-Brillouin (WKB) approximation, that was
pioneered in Refs.~\cite{Kubo,Hu,Peters,DykmanPRE100},
and was applied more recently to population extinction, in particular, for stochastic
two-population systems \cite{DykmanPRL101,MeersonPRE77,MS09,
  KhasinPRL103,MeersonPRE81,LM2011,BM11,GM12,KMKS}. The WKB ansatz for $\Pi_{m,n}$
has the form
\begin{equation}
\label{WKB}
\Pi_{m,n}= e^{-KS(x,y)},
\end{equation}
where $x=m/K$ and $y=n/K$ are treated as continuous
variables.
Substituting Eq.~\eqref{WKB} into
\eqref{rate1} and assuming $K\gg 1$, gives, to lowest order in
$1/K$
\begin{eqnarray}
\label{rate2}
R_A\sim e^{-KS_A}\,,\qquad\qquad
R_B\sim e^{-KS_B}\,,
\end{eqnarray}
where $S_A=S(0,g)$ and $S_B=S(1,0)$, (see
Refs.~\cite{DykmanPRL101,MeersonPRE77,MS09,GM12}). 
Thus as $K\gg 1$, the eventual extinction probabilities in \eqref{extinction}
simply become (up to pre-exponential factors that depend on $K$), as
\begin{equation}
\mathcal{P}_A(t\!=\!\infty)=1\!-\!\mathcal{P}_B(t\!=\!\infty)
\simeq \frac{e^{-KS_A}}{e^{-KS_A}\!+\!e^{-KS_B}}~.
\label{Eprob}
\end{equation}

To determine the extinction probabilities explicitly, we therefore need $S_A$
and $S_B$.  To this end, we substitute the WKB ansatz~(\ref{WKB}) in
Eq.~(\ref{eigen}) and Taylor expand $S(x,y)$ to lowest order in $1/K$.
After some algebra, we obtain an effective Hamilton-Jacobi equation
$H(x,y,\partial_x S, \partial_y S)=-\mathcal{R}$, with the Hamiltonian
\begin{align}
\label{H}
\begin{split}
H(x,y,p_x,p_y) &= x(e^{p_x}\!-\!1)+gy(e^{p_y}\!-\!1)\\
& +\!\frac{x^2}{2}(e^{-2p_x}\!-\!1)\! +\!\frac{y^2}{2}(e^{-2p_y}\!-\!1) \\
& +\epsilon xy(e^{-p_x}\!-\!1)\!+\!\alpha\epsilon xy(e^{-p_y}\!-\!1) \,.
\end{split}
\end{align}
Here $p_x=\partial_x S$ and $p_y=\partial_y S$ are the canonical momenta that
are conjugate to the ``coordinates" $x$ and $y$.  Correspondingly, $S(x,y)$
is the classical action of the system.

Since we expect that $-\mathcal{R}$ (which now has the meaning of energy in
this Hamiltonian system) is expected to be exponentially small, we set it to
zero.  The Hamiltonian equations of motion $\dot{x}=\partial H/\partial p_x$,
$\dot{p}_x=-\partial H/\partial x$, etc., have six finite zero-energy fixed
points, and three more fixed points where one or both momenta are at minus
infinity.  
Only three of the fixed points, however, turn out to be relevant for
answering our question about which species typically goes extinct first.
These are:
\begin{align}
\begin{split}
  F_\phi&= \left(\frac{1-g\epsilon}{1-\alpha\epsilon^2},\frac{g-\alpha\epsilon}{1-\alpha\epsilon^2}, 0,0\right)\,, \\
  F_A&=\left( 0,g,\ln(g\epsilon),0\right)\,, \\
  F_B&=\left( 1,0,0,\ln(\alpha\epsilon/g)\right)\,.
\end{split}
\end{align}
A straightforward way to determine $S_A$ and $S_B$ is by calculating the
action along the \emph{activation trajectories}.  These are zero-energy, but
non-zero-momentum trajectories of the Hamiltonian system (\ref{H}) that go
from $F_\phi$ to $F_A$, and from $F_\phi$ to $F_B$, respectively.  These
actions are
\begin{equation}
S_{A}=\int_{F_\phi}^{F_{A}} (p_xdx+p_ydy)\,,
\label{action}
\end{equation}
and similarly for $S_{B}$. In general, these activation
trajectories---separatrices, or instantons---cannot be calculated
analytically because of the lack of an integral of motion that is independent
of the energy.  However, for $\epsilon\ll 1$ a perturbative solution for these
trajectories is possible.

As a preliminary, we outline how to calculate the action for the special case
$\epsilon=0$, which corresponds to two uncoupled species.  Here the
zero-energy activation trajectories can be easily found.  For the
$F_\phi\rightarrow F_A$ separatrix, the B species is unaffected by A
extinction so $(y,p_y)$, which correspond to the coordinates of the Bs,
remains constant throughout the evolution.  As a result, a parametric form of
the $F_\phi\rightarrow F_A$ separatrix is
\begin{subequations}
\label{trajectories}
\begin{eqnarray}
x=\frac{2e^{2p_x}}{e^{p_x}\!+\!1},
\end{eqnarray}
with $y=g$ and $p_y=0$ throughout.  Similarly, for the $F_\phi\rightarrow
F_B$ separatrix one obtains
\begin{eqnarray}
y=\frac{2g e^{2p_y}}{e^{p_y}\!+\!1},
\end{eqnarray}
\end{subequations}
with $x=1$ and $p_x=0$ throughout.  Substituting the trajectories given
in~\eqref{trajectories} into~\eqref{action} and performing the integration by
parts gives $S_A=2(1-\ln 2)$ and $S_B=2 g(1-\ln 2)$ \cite{EK,KS07}.

For weak interspecies competition ($\epsilon\ll 1$), we can calculate the
corrections to the actions to first order in $\epsilon$.  For this purpose,
we split the Hamiltonian (\ref{H}) into unperturbed and perturbed parts,
$H=H_0+\epsilon H_1$, and similarly expand the action as
$S=S_0+\epsilon S_1+\dots$.  Following
\cite{DykmanPRL101,AKM08,KhasinPRL103}, the correction to the action is
$S_1=\int_{-\infty}^{\infty} H_1[x(t),y(t),p_x(t),p_y(t)] dt$, where
the integral is evaluated along the {\it unperturbed} trajectories given by
Eqs.~\eqref{trajectories}.  Performing this integral for $S_1$ yields the
corrected actions
\begin{align}
\begin{split}
\label{correction}
S_A&=2(1-\ln 2) -\epsilon (2g\ln 2) \,,\\
S_B&=2 g(1-\ln2) -\epsilon(2\alpha\ln 2)\,.
\end{split}
\end{align}

\begin{figure}[ht]
\center{
\includegraphics[width=0.36\textwidth]{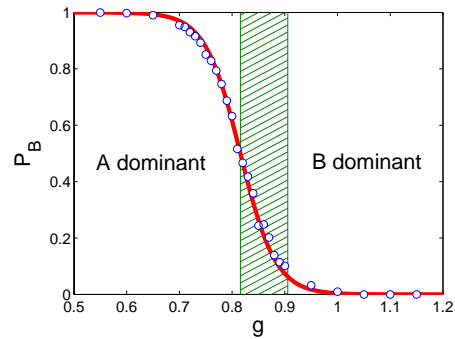}
\caption{Probability that species B first becomes extinct as a function of
  growth rate asymmetry $g$ for $\alpha=0$, $\epsilon=0.1$, $K=40$.  The
  curve is the prediction of Eq.~\eqref{Eprob}, with actions given by
  \eqref{correction}, while circles are simulation results.  In the hatched
  region, the quasi-stationary population of B is less than that of A, but Bs
  are less likely to become extinct first. }
\label{extPsmallE}
}
\end{figure}

Equations~\eqref{rate2} and~\eqref{correction} give the analytic expression
for the extinction probabilities of each species for weak interspecies
competition.  Using Eq.~\eqref{correction} and imposing the condition
$S_A=S_B$ from Eq.~\eqref{Eprob}, we obtain the following condition for equal
extinction probability for both species:
\begin{equation}
\label{halfProb}
g=\frac{1+\alpha\,c\,\epsilon}{1+c\,\epsilon}~,
\end{equation}
where $c=\big[(\ln 2)^{-1}\!-\!1\big]^{-1}$.  The predictions of
Eqs.~\eqref{Eprob} and \eqref{correction} are in good agreement with our
simulation results (Fig.~\ref{extPsmallE}).

\begin{figure}[htb]
\center{
\includegraphics[width=0.38\textwidth]{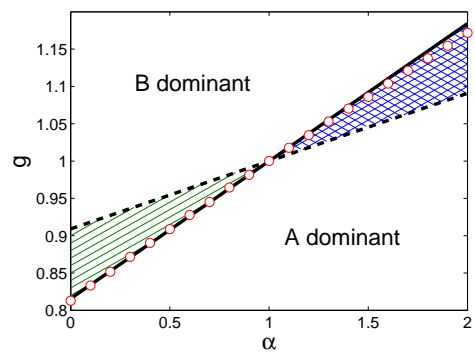}
\caption{ Phase diagram for $\epsilon=0.1$ showing loci of equal
  quasi-stationary population sizes (dashed) and equal extinction probabilities
  from Eq.~\eqref{halfProb} (solid).  Circles indicate simulation results.
  In the hatched region, As are more numerous in the quasi-stationary state
  but are more likely to become extinct first.  In the cross-hatched region
  Bs are more numerous but are more likely to become extinct first.
}
\label{phase}
}
\end{figure}

\medskip {\it Phase Diagram:} Comparing Eqs.~\eqref{equalsize} and
\eqref{halfProb}, one sees that there is a sizable region in the $\alpha$-$g$
parameter space where one species has a \emph{smaller} quasi-stationary
population and yet an (exponentially) \emph{smaller} probability to first
become extinct.  As an illustration, Fig.~\ref{extPsmallE} shows the
probability for B to become extinct first for fixed $\alpha$ and $\epsilon$.
We also produced analogous curves as Fig.~\ref{extPsmallE} at many values of
$\alpha$.  From the value of $g$ at which the extinction probabilities are
equal, we infer the phase diagram shown in Fig.~\ref{phase}.  Simulations at
larger values of $\epsilon$ yield the same qualitative phase diagram.

\medskip {\it Conclusion:} In two-species competition, interspecies
competitive asymmetry leads to the unexpected phenomenon of ``survival of the
scarcer''.  The very same elemental reactions that lead to a disadvantage in
the quasi-stationary population size of one species within a deterministic
mean-field theory, may also give this species a great advantage in its
long-term survival when fluctuation effects are properly accounted for.

\smallskip AG and SR gratefully acknowledge NSF grant DMR-0906504 and
DMR-1205797 for partial financial support.  BM was partially supported by the
Israel Science Foundation (Grant No.\ 408/08), by the US-Israel Binational
Science Foundation (Grant No.\ 2008075), and by the Condensed Matter Theory
Visitors Program in the Boston University Physics Department.

\end{document}